\documentclass[a4paper,12pt,english]{article}
\usepackage{amsfonts}
\usepackage{graphicx}
\usepackage{amsmath}
\usepackage{color}

\newcommand{\be}{\begin{equation}}
\newcommand{\ee}{\end{equation}}
\input{tcilatex}

\begin{document}

\begin{titlepage}
\begin{center}

\noindent{{\LARGE{On the description of surface operators in ${\mathcal N}=2^*$ SYM}}}

\smallskip
\smallskip

\smallskip
\smallskip
\smallskip
\smallskip
\smallskip
\smallskip
\noindent{\large{Juan Pablo Babaro and Gaston Giribet}}

\smallskip
\smallskip

\end{center}
\smallskip
\smallskip
\centerline{Departamento de F\'{\i}sica, Universidad de Buenos Aires FCEN-UBA and IFIBA-CONICET}
\centerline{{\it Ciudad Universitaria, Pabell\'on 1, 1428, Buenos Aires, Argentina.}}
\smallskip
\smallskip

\bigskip

\bigskip

\bigskip

\bigskip

\begin{abstract}
In Ref. \cite{AT}, Alday and Tachikawa observed that the Nekrasov partition function of ${\mathcal N}=2$
$SU(2)$ superconformal gauge theories in the presence of fundamental surface operators can be associated
to conformal blocks of a 2D CFT with affine $sl(2)$ symmetry. This can be interpreted as the insertion of
a fundamental surface operator changing the conformal symmetry from the Virasoro symmetry discovered in
\cite{AGT1} to the affine Kac-Moody symmetry.
A natural question arises as to how such a 2D CFT description can be extended to the case of
non-fundamental surface operators.
Motivated by this question, we review the results of Refs. \cite{Schomerus} and \cite{Ribault-family} and put them together to suggest a way to address the problem:
It follows from this analysis that the expectation value of a non-fundamental surface operator in the $SU(2)$ ${\mathcal N}=2^*$ super Yang-Mills theory
would be in correspondence with the expectation value of a single vertex operator in a two-dimensional CFT with reduced affine symmetry and whose central charge is parameterized by the integer number that
labels the type of singularity of the surface operator.
\end{abstract}

\end{titlepage}

\newpage

\section{Introduction}

Of fundamental importance in theoretical physics is the question about
non-perturbative effects in Yang-Mills theory (YM). In the last two decades,
there has been important progress in this area, mainly due to our current
understanding of the supersymmetric extensions of the theory. In the last
few years, one of the most promising advances in the direction of
understanding non-perturbative effects of supersymmetric YM theories has
been the observation, due to Alday, Gaiotto, and Tachikawa \cite{AGT1}, that
the Nekrasov partition functions \cite{Nekrasov} of certain class of $%
\mathcal{N}=2$ superconformal $SU(2)$ quiver theories in four dimensions are
given by the conformal blocks of Liouville field theory. According to this,
the full partition function of such gauge theories, meaning the partition
function including instanton corrections, would be abstrusely encoded in the
building blocks of a relatively well undertsood two-dimensional conformal
field theory (CFT).

Specifically, Alday-Gaiotto-Tachikawa conjecture (AGT) states that the $n$%
-point conformal blocks of Liouville field theory formulated on an $n$%
-punctured genus-$g$ Riemann surface $\mathcal{C}_{g,n}$ give the Nekrasov
partition function of the Gaiotto's quiver theory $\mathcal{T}_{g,n}$ that
is constructed, as in \cite{Gaiotto}, by compactifying the six-dimensional
(2,0) theory of the A$_{1}$ type on $\mathcal{C}_{g,n}$. This provides a
very interesting correspondence between 2D conformal field theories and 4D
superconformal gauge theories.

Soon after the paper \cite{AGT1} appeared, the extension of the 2D/4D
correspondence to the cases in which both loop and surface operators are
incorporated on the gauge theory side was proposed \cite{AGT2,AGT3}. In this
generalized picture, not only the partition function, but also the
expectation values of defects in the 4D theory happen to be described by
Liouville correlation functions. It turns out that, to the insertion of a
defect on the 4D gauge theory side, it corresponds the insertion of a
degenerate Liouville field in the 2D CFT side.

More recently, it has been observed that the 2D CFT\ description of
expectation values of defects in the gauge theory is naturally realized in
terms of CFTs with affine symmetry \cite{AT}. For the case of $SU(2)$ gauge
theories, this involves CFT with $\widehat{sl}(2)_{k}$ affine Kac-Moody
symmetry, with $k=\varepsilon _{1}/\varepsilon _{2}+2$, with $\varepsilon
_{1,2}$ being the Nekrasov's deformation parameters \cite{Nekrasov}. In some
sense, it is adequate to say that, while Liouville field theory stands as
the convenient language to represent the $\mathcal{N}=2$ gauge theory
partition function, the expectation values of defects in such theories are
more conveniently described by conformal blocks of 2D CFTs with affine
symmetry; at least it seems to be the case for the simplest defects. One of
the motivations of this paper is to propose a way of extending such affine
CFT realization to the case of non-fundamental surface operators.

\section{Surface operators and CFT}

Our current understanding of the problem is that non-fundamental surface
operators whose vacua are labeled by an integer number $m\geq 1$ exist in
these $SU(2)$ $\mathcal{N}=2$ gauge theories, and the expectation value of
such a surface operator would admit a 2D CFT description in terms of a
Liouville correlation function with the additional insertion of a degenerate
field of conformal dimension $h=-(m/2)(1+b^{-2}(1+m/2))$, namely the vertex $%
e^{-m\varphi (x)/b}$. Inconveniently, a purely gauge theory description of
the surface operators that correspond to $m>1$ is still missing, and without
a complete description of defects from the gauge theory point of view it is
worthwhile studying the problem from different perspectives. Here, with the
aim of contributing to the study of non-fundamental surface operators in the
4D $\mathcal{N}=2$ SCFTs, we will draw the attention to a yet unexplored
CFTs tool developed in Ref. \cite{Ribault-family}. We will focus our
attention on the particular case of $SU(2)$ $\mathcal{N}=2^{\ast }$ SYM.
Invoking the result of \cite{Ribault-family}, or more precisely its
genus-one generalization, we will argue that the expectation value of a
non-fundamental surface operator (labeled by an integer $m$) in the $%
\mathcal{N}=2^{\ast }$ theory is given by the expectation value of a single
vertex operator in a 2D CFT which has central charge $%
c_{(b,m)}=3+6(b^{-1}+(1-m)b)^{2}$, with $b^{2}=\varepsilon _{1}/\varepsilon
_{2}$.

Our line of argument is the following:\ According to the results of Ref. 
\cite{AGT2}, the expectation value of a surface operator in $SU(2)$ $%
\mathcal{N}=2^{\ast }$ SYM\ is associated to the Liouville 2-point function
\ $\left\langle e^{2\alpha \varphi }e^{-m\varphi /b}\right\rangle $ on the
torus. On the other hand, a genus-one extension of the result of \cite%
{Ribault-family} permits to express such Liouville 2-point function in terms
of the expectation value of a single vertex operator $\left\langle \Phi
_{h}\right\rangle $ in a 2D CFT with central charge $c_{(b,m)}$ given above.
In the case $m=1$, which corresponds to fundamental surface operators, the
2D CFT is identified with the $\widehat{sl}(2)_{k}$ affine theory with level 
$k=2+1/b^{2}$, suggesting a possible connection with the analysis of \cite%
{AT}. In the case $m>1$, in contrast, such 2D CFT exhibits only part of the
affine symmetry, being this part generated by the Borel subalgebra of $%
\widehat{sl}(2)_{k^{\prime }}$ with $k^{\prime }=2+m^{2}/b^{2}$.

As said, the specific CFTs we will consider are those that were introduced
by Ribault in Ref. \cite{Ribault-family}. It was shown in \cite%
{Ribault-family} that the genus-$0$ ($2n-2$)-point Liouville correlation
functions that involve $n-2$ degenerate fields $e^{-m\varphi /b}$ with $m\in 
\mathbb{Z}_{\geq 0}$ provide the $n$-point correlation functions of a
non-rational CFT with central charge $c=3+6(b^{-1}+(1-m)b)^{2}$. These CFTs,
if they actually exist for generic $m\in \mathbb{Z}_{>1}$, would coincide
with the $\widehat{sl}(2)_{k}$ WZW theory for $m=1$, and with Liouville
theory itself for $m=0$. The point we want to make here is that other
members of this $m$-labeled family of CFTs may have application to describe
obervables in 4D gauge theories as well. Here we will be concerned with the $%
\mathcal{N}=2^{\ast }$ SYM theory, and then, according to \cite{AGT2}, this
corresponds to the 2D CFT being formulated on the torus. Then, we first need
to solve a preliminary problem: we need to extend the construction of \cite%
{Ribault-family} to genus one, $g=1$. This is basically the result of this
paper:\ in what follows we will show that, as probably expected, the torus
Liouville $2n$-point functions that involve $n$ degenerate fields $%
e^{-\varphi m/b}$ and $n$ generic primary fields $e^{-2\alpha _{i}\varphi }$
with $\alpha _{i}\in \mathbb{C}$ actually coincide with the torus $n$-point
function of the $m^{\text{th}}$ member of the family of CFTs proposed in 
\cite{Ribault-family}. Proving so follows from straightforwardly adapting
the path integral techniques developed in Ref. \cite{Schomerus} to the $m>1$
case. From the CFT point of view, this result is interesting in its own
right as it provides further evidence of the consistency of the theories
proposed in \cite{Ribault-family}.

We begin in Sections III and IV by reviewing the theories introduced in \cite%
{Ribault-family} formulated on the genus-$0$ surface. In Section V, we
extend the result of \cite{Ribault-family} to genus-$1$ by using the
techniques developed in \cite{Schomerus}; this represents a trivial
extension of the results therein. Section VI contains the conclusions.

\section{A family of conformal field theories}

Let us introduce the family of CFTs we will be concerned with. Each member
of the family, each CFT, depends on two parameters, $m$ and $b$. Here we
will consider $m\in \mathbb{Z}_{\geq 0}$ and $b\in \mathbb{R}_{>0}$. The
action can be written as \cite{Ribault-family} 
\begin{equation}
S_{(m,b)}[\phi ,\beta ,\gamma ]\ =\ \frac{1}{2\pi }\int d^{2}z\,\left(
\partial \phi \bar{\partial}\phi +\beta \bar{\partial}\gamma +\bar{\beta}%
\partial \bar{\gamma}+\frac{Q_{(m,b)}}{4}\mathcal{R}\phi +b^{2}(-\beta \bar{%
\beta})^{m}\,e^{2b\phi }\right) ,  \label{accion}
\end{equation}%
where the background charge takes the value 
\begin{equation}
Q_{(m,b)}\ =\ b+\frac{1-m}{b}.  \label{uf}
\end{equation}

Let us call $Y_{m,b}$ the theory defined by the action (\ref{accion}). In (%
\ref{accion}) $\mathcal{R}$ represents the scalar curvature of the surface
on which the theory is defined. From this action we observe that $Y_{0,b}$
corresponds to Liouville field theory with central carge $c=1+6(b+1/b)^{2}$
coupled to a free $\beta $-$\gamma $ ghost system. On the other hand, theory 
$Y_{1,b}$ corresponds to the $\mathbb{H}_{3}^{+}=SL(2,\mathbb{C})/SU(2)$
WZNW theory with level $k=b^{-2}+2$ expressed in the Wakimoto free-field
representation \cite{waki}. Theory $Y_{b^{2},1/b}$ also corresponds to the $%
\mathbb{H}_{3}^{+}$ WZNW theory with the Langlands dual level $k^{\text{L}%
}=b^{+2}+2$ \cite{Giribet-Nicolas, TeschnerLanglands}.

The stress-tensor associated to the free theory defined by action (\ref%
{accion}) is given by 
\begin{equation}
T(z)=-\beta (z)\partial \gamma (z)-(\partial \phi
(z))^{2}+Q_{(m,b)}\,\partial ^{2}\phi (z)  \label{T}
\end{equation}%
and by its anti-holomorphic counterpart $\overline{T}_{(\bar{z})}$. This
gives the central charge of the theory 
\begin{equation}
c_{(m,b)}\ =\ 1+6Q_{(m,b)}^{2}.  \label{ccc}
\end{equation}

Last term in the (\ref{accion}) represents a marginal operator with respect
to the stress-tensor (\ref{T}) as it can easily be verified by using the
free field propagators.

As mentioned, the theory with $m=1$ corresponds to the WZNW model, which
exhibits $\widehat{sl}(2)_{k}\times \widehat{sl}(2)_{k}$ affine Kac-Moody
symmetry. This symmetry is generated by the currents 
\begin{equation}
J^{-}(z)=\beta (z),\ \ \ \ \ \ \ \ J^{3}(z)=\beta (z)\gamma
(z)+b^{-1}\partial \phi (z),  \label{J3}
\end{equation}%
and 
\begin{equation}
J^{+}(z)=\beta (z)\gamma ^{2}(z)+2b^{-1}\gamma (z)\partial \phi (z)-\left(
b^{-2}+2\right) \partial \gamma (z),  \label{J5}
\end{equation}%
together with the anti-holomorphic counterparts $\bar{J}^{3}(\bar{z}),$ $%
\bar{J}^{\pm }(\bar{z})$, where $b^{-2}=k-2$. In contrast, for the theory
with $m\neq 1$ only a sub-algebra of (\ref{J3})-(\ref{J5}) survives and the
theory exhibits a remaining symmetry under the Borel subalgebra of $\widehat{%
sl}(2)_{k^{\prime }}$ with modified level $k^{\prime }=2+m^{2}/b^{2}$
generated by the currents%
\begin{equation}
J^{-}(z)=\beta (z),\qquad \qquad J^{3}(z)=\beta (z)\gamma (z)+\frac{m}{b}%
\partial \phi (z)  \label{14}
\end{equation}%
and the anti-holomorphic analogues. Currents (\ref{14}) obey the following
operator product expansion 
\begin{equation}
J^{-}(z)J^{3}(w)\ \simeq \ \frac{J^{-}(w)}{(z-w)}+...\qquad
J^{3}(z)J^{3}(w)\ \simeq \ -\frac{(1+m^{2}b^{-2}/2)}{(z-w)^{2}}+...\qquad
J^{-}(z)J^{-}(w)\ \simeq \ ...  \label{borel}
\end{equation}%
where the ellipses stand for regular terms that are omitted. This realizes
the Lie brackets%
\begin{equation*}
\lbrack J_{p}^{-},J_{q}^{3}]=J_{p+q,0}^{-},\qquad \lbrack
J_{p}^{3},J_{q}^{3}]=-p\frac{k^{\prime }}{2}\delta _{p+q,0},\qquad \lbrack
J_{p}^{-},J_{q}^{-}]=0,\qquad
\end{equation*}
for the modes defined as usual, 
\begin{equation}
J_{p}^{3}=\frac{1}{2\pi i}\int dz\ z^{-p-1}J^{3}(z),\ \ \ \ \ \ \ J_{p}^{-}=%
\frac{1}{2\pi i}\int dz\ z^{-p-1}J^{-}(z).
\end{equation}

The spectrum of the theory would consist of fields that are primary with
respect to currents (\ref{14}). Vertex operators creating such states are of
the form 
\begin{equation}
\Phi _{h}(\mu |z)=\,|\mu |^{2m(j+1)}\,e^{\mu \gamma (z)-\bar{\mu}\bar{\gamma}%
(\bar{z})}\,e^{2b(j+1)\phi (z,\bar{z})}  \label{CuatrO}
\end{equation}%
whose holomorphic and anti-holomorphic conformal dimensions are given by $%
h_{j}\ =\ \bar{h}_{j}\ =(-b^{2}j+1-m)(j+1)$. In (\ref{CuatrO}), $\mu $ is a
complex variables and $j$ is an index that can be thought of as the momentum
of the field. The spectrum of normalizable states of the theory would be
ultimately determined by the values that $j$ takes. Fields $\Phi _{h}(\mu
|z) $ should also include an overall factor $|\omega _{(z,\bar{z}%
)}|^{2h_{j}} $ with $\omega _{(z,\bar{z})}$ being the Weyl factor of the
two-dimensional metric $ds^{2}=|\omega _{(z,\bar{z})}|^{2}dzd\bar{z}$ on the
surface. Such dependence of the conformal factor is required for $\Phi
_{j}(\mu |z)$ to transform as a primary ($h_{j},\overline{h}_{j}$)-dimension
operator. For short, we will set $\omega _{(z,\bar{z})}=1$ in formulae below.

\section{Genus-zero correlation functions}

Let us begin by defining the theory defined by (\ref{accion}) on the sphere
topology. More precisely, let us calculate the genus-zero correlation
functions 
\begin{equation}
\Omega _{(m,b)}^{g=0,n}\left( \mu _{\nu }|z_{\nu }\right) \equiv
\left\langle \prod_{\nu =1}^{n}\Phi _{h_{\nu }}(\mu _{\nu }|z_{\nu
})\right\rangle =\int \mathcal{D}\phi \mathcal{D}^{2}\beta \mathcal{D}%
^{2}\gamma \,e^{-S_{(m,b)}[\phi ,\beta ,\gamma ]}\prod_{\nu =1}^{n}\Phi
_{h_{\nu }}(\mu _{\nu }|z_{\nu }),
\end{equation}%
where the expectation value is understood as the correlation function of
primary operators (\ref{CuatrO}) in the theory $Y_{m,b}$ formulated on the $%
n $-puncture sphere $\mathcal{C}_{0,n}$.

Functional integration over the fields $\gamma $ and $\bar{\gamma}$ yields $%
\delta $-functions that fix the conditions%
\begin{equation}
\bar{\partial}\beta (w)=2\pi \sum_{\nu =1}^{n}\mu _{\nu }\,\delta
^{2}(w-z_{\nu }),\ \ \ \ \ \partial \bar{\beta}(\bar{w})=-2\pi \sum_{\nu
=1}^{n}\bar{\mu}_{\nu }\,\delta ^{2}(\bar{w}-\bar{z}_{\nu }).  \label{doce}
\end{equation}%
These equations have solution only if $\sum_{\nu =1}^{n}\mu _{\nu }=0$.
Having in mind that $\bar{\partial}\left( 1/z\right) \ =\ \partial \left( 1/%
\bar{z}\right) \ =\ 2\pi \,\delta ^{2}(z)$, we write the most general
solution to (\ref{doce}) in the form \cite{Schomerus}%
\begin{equation}
\beta (w)=\sum_{\nu =1}^{n}\mu _{\nu }(w-z_{\nu })^{-1}=u\frac{%
\prod_{i=1}^{n-2}(w-y_{i})}{\prod_{j=1}^{n}(w-z_{\nu })},
\end{equation}%
with the following relation%
\begin{equation}
\mu _{\nu }\ =\ u\,\frac{\prod_{i=1}^{n-2}(z_{\nu }-y_{i})}{\prod_{\mu \neq
\nu }^{n}(z_{\nu }-z_{\mu })},\qquad \sum_{\nu =1}^{n}\mu _{\nu }=0,
\label{Estas}
\end{equation}%
where a new variable $u$ and $n-1$ variables $y_{i}$ have been introduced. (%
\ref{Estas}) are $n-1$ equations that permit to express the $n-1$
independent variables $\mu _{\nu }$ in terms of the $n-2$ variables $y_{i}$
and the variable $u$. This permits to integrate over $\beta $ and $\bar{\beta%
}$ and obtain

\begin{equation}
\Omega _{(m,b)}^{g=0,n}\left( \mu _{\nu }|z_{\nu }\right) =\int \mathcal{D}%
\phi \,\ e^{-S_{\text{eff}}[\phi ,\mathcal{X}_{0}]}\,\prod_{\nu =1}^{n}|\mu
_{\nu }|^{2m(j_{\nu }+1)}\,e^{2b(j_{\nu }+1)\phi (z_{\nu })}.
\end{equation}%
with the effective action

\begin{equation}
S_{\text{eff}}[\phi ,\mathcal{X}_{0}]=\frac{1}{2\pi }\int d^{2}z\left(
\partial \phi \bar{\partial}\phi +\frac{Q_{(m,b)}}{4}\mathcal{R}\phi
\,+b^{2}|u|^{2m}\left\vert \mathcal{X}_{0}\right\vert ^{2m}e^{2b\phi }\right)
\label{Eef}
\end{equation}%
with%
\begin{equation}
\mathcal{X}_{0}(y_{i},z_{\nu };w)\equiv \frac{\prod_{i=1}^{n-2}(w-y_{i})}{%
\prod_{\nu =1}^{n}(w-z_{\nu })};
\end{equation}%
see \cite{Schomerus} for details.

The next step is to massage expression (\ref{Eef}) to bring it into its
Liouville form. To achieve so, one first performs the shifting $\phi (w,\bar{%
w})\rightarrow \phi (w,\bar{w})-(m/b)\,\log |u|$, and arrives to 
\begin{equation*}
\Omega _{(m,b)}^{g=0,n}\left( \mu _{\nu }|z_{\nu }\right) =|u|^{2m\left(
1+(1-m)/b^{2}\right) }\int \mathcal{D}\phi \,\ e^{-S_{\text{eff}}[\phi ,%
\mathcal{X}_{1}]}\prod_{\nu =1}^{n}|\mu _{\nu }|^{2m(j_{\nu
}+1)}\,|u|^{-2m(j_{\nu }+1)}e^{2b(j_{\nu }+1)\phi (z_{\nu })}.
\end{equation*}%
Then, defining the new variable%
\begin{equation}
\phi \equiv \varphi \,-\,\frac{m}{2b}(\sum_{i=1}^{n-2}\log
|w-y_{i}|^{2}-\sum_{\nu =1}^{n}\log |w-z_{\nu }|^{2}-\log |\omega _{(w,%
\overline{w})}|^{2}),
\end{equation}%
and taking into account the powers of the conformal factor $\log |\omega
_{(z,\overline{z})}|^{2}$ generated in the regularized coincident limit $%
w\rightarrow z_{\nu }$, one finds that the background charge changes as $%
Q_{(m,b)}\rightarrow Q_{(0,b)}\,=\,b+1/b$. The latter corresponds to the
Liouville background charge.

Finally, setting the overall factor $|u|^{2m(1+(1-m)/b^{2})}$ to one, one
finds the expression obtained in \cite{Ribault-family}; namely 
\begin{equation}
\Omega _{(m,b)}^{g=0,n}\left( \mu _{\nu }|z_{\nu }\right) =\left\langle
\prod_{\nu =1}^{n}e^{i\frac{m}{b}X(z_{\nu })}\prod_{i=1}^{n-2}e^{-i\frac{m}{b%
}X(y_{i})}\right\rangle _{X}\left\langle \prod_{\nu =1}^{n}V_{\alpha _{\nu
}}(z_{\nu })\prod_{i=1}^{n-2}V_{-\frac{m}{2b}}(y_{i})\right\rangle _{\text{L}%
}  \label{RTrelation}
\end{equation}%
which is subject to conditions (\ref{Estas}), in particular to the condition 
$\sum_{\nu =1}^{n}\mu _{\nu }=0$. On the right hand side of (\ref{RTrelation}%
), the Liouville correlation functions are given by 
\begin{equation}
\left\langle \prod_{\nu =1}^{n}V_{\alpha _{\nu }}(z_{\nu
})\prod_{i=1}^{n-2}V_{-\frac{m}{2b}}(y_{i})\right\rangle _{\text{L}}\ =\
\int \mathcal{D}\varphi \,e^{-S_{\text{L}}[\varphi ]}\prod_{\nu
=1}^{n}e^{2\alpha _{\nu }\varphi (z_{\nu })}\prod_{i=1}^{n-2}e^{-\frac{m}{b}%
\varphi (y_{i})}
\end{equation}%
with the Liouville action 
\begin{equation*}
S_{\text{L}}[\varphi ]\ =\ \frac{1}{2\pi }\int d^{2}z\,(\partial \varphi 
\bar{\partial}\varphi \,+\,\frac{1}{4}(b+b^{-1})\sqrt{g}\mathcal{R}\varphi
\,+\,b^{2}e^{2b\varphi })
\end{equation*}%
and with momenta $\alpha _{\nu }\ =\ b(j_{\nu }+1+b^{-2}/2)$. The overall
factor is of the form%
\begin{equation}
\left\langle \prod_{\nu =1}^{n}e^{i\frac{m}{b}X(z_{\nu
})}\prod_{i=1}^{n-2}e^{-i\frac{m}{b}X(y_{i})}\right\rangle _{X}=\prod_{\mu
<\nu }^{n}|z_{\mu }-z_{\nu
}|^{m^{2}b^{-2}}\prod_{i<j}^{n}|y_{i}-y_{j}|^{m^{2}b^{-2}}\prod_{\mu
=1}^{n}\prod_{i=1}^{n-2}|z_{\mu }-y_{i}|^{-m^{2}b^{-2}},  \label{RTrelation2}
\end{equation}%
which can be interpreted as the correlation function of a free boson $X(z)$
with non-trivial background charge $\widehat{Q}=im/b$. That is, equation (%
\ref{RTrelation}) can be thought of as expressing the equivalence between $n$%
-point correlation functions of the theory defined by action (\ref{accion})
and ($2n-2$)-point correlation functions of a theory composed by Liouville
theory times a CFT with central charge $c=1-6m^{2}/b^{2}$. Expression (\ref%
{RTrelation}) generalizes the relation between between the $\mathbb{H}%
_{3}^{+}$ WZNW theory and Liouville field theory derived by Stoyanovsky \cite%
{Stoyanovsky} and by Ribault and Teschner \cite{Ribault-Teschner}, which is
reobtained by replacing $m=1$ in the formulae above. Here we have reviewed
the derivation of (\ref{RTrelation}) given by Hikida and Schomerus in Ref. 
\cite{Schomerus}, which, as we will see in the next Section, can be
generalized to genus-one.

\section{Genus-one correlation functions}

Now, let us consider the theories (\ref{accion}) on the genus-one surface.
We will follow the analysis of Ref. \cite{Schomerus}, adapting it to the
case $m>1$.

As usual, the torus is represented by the complex plane with periodic
conditions under translations $w\rightarrow w+1$ and $w\rightarrow w+\tau $.
The complex variable $\tau =\tau _{1}+i\tau _{2}$ is the modular parameter
of the torus. To fully parameterize the consistent boundary conditions, it
is also necessary to introduce an additional parameter $\lambda $ which
amounts to consider twisted periodicity conditions 
\begin{equation}
\beta (w+p+q\tau )\ =\ e^{2\pi iq\lambda }\beta (w),\qquad \gamma (w+p+q\tau
)\ =\ e^{-2\pi iq\lambda }\gamma (w),  \label{gammabc}
\end{equation}%
and%
\begin{equation}
\phi (w+p+q\tau \,,\,\bar{w}+p+q\bar{\tau})\ =\ \phi (w,\bar{w})+\frac{2\pi
mq\text{Im}\lambda }{b},  \label{phi bc}
\end{equation}%
where $\text{Im}\lambda =\lambda _{2}$ stands for the imaginary part of the
twist parameter $\lambda =\lambda _{1}+i\lambda _{2}$, and $p$ and $q$ are
two arbitrary integer numbers that parameterize the steps on the lattice.
The possibility of choosing conditions (\ref{gammabc})-(\ref{phi bc}), even
when they yield multivalued fields for $\lambda \neq 0$, comes from the fact
that action (\ref{accion}) does remain univalued. For $\lambda =0$,
untwisted boundary conditions are recovered. For $m=0$ the field $\phi $
must be periodic; it acquires more freedom in the case $m\neq 0$ and such is
parameterized by $\lambda $, which labels different twist sectors. For $m=1$%
, $\lambda $ is identified with the Benard parameter that appears in the
Bernard-Knizhnik-Zamolodchikov modular differential equation; see \cite%
{Schomerus} for a detailed discussion. For $m>1$, as long as $m\neq b^{2}$,
the theory does not exhibit the full $\widehat{sl}(2)_{k}$ affine symmetry;
nevertheless, $\lambda $ may still be introduced.

The next step, before integrating over $\beta $ and $\gamma $, is to
decompose the field $\phi $ into its solitonic zero-mode part%
\begin{equation}
\phi _{\text{c}}(w,\bar{w})\ =\ \frac{2\pi m}{b}\frac{\text{Im}(\lambda )%
\text{Im}(w)}{\text{Im}(\tau )}.  \label{d8}
\end{equation}
and the fluctuations $\phi _{\text{f}}$; namely $\phi \ (w,\bar{w})=\phi _{%
\text{c}}(w,\bar{w})+\phi _{\text{f}}(w,\bar{w})$. Solitonic configuration (%
\ref{d8}), together with $\beta =0$ and $\gamma =0$, represents the only
solution to the classical equations of motion coming from the action (\ref%
{accion}) that satisfies the required periodic boundary conditions. The
piece $\phi _{\text{f}}$ is periodic under $w\rightarrow w+1$ and $%
w\rightarrow w+\tau $. We are nowon the torus, so only the fluctuations $%
\phi _{\text{f}}$ couple to the scalar curvature in the linear dilaton term.
Although one considers the flat metric on the genus-one surface, this term
is ultimately important to keep track of the background charge contribution
when expressing the final result in terms of the Liouville field theory
analogue. It can be restored wherever needed by writing the action (\ref%
{accion}).

Then, we are ready to compute the correlation functions. These are defined by%
\begin{equation}
\Omega _{(m,b)}^{g=1,n}=\left\langle \prod_{\nu =1}^{n}\Phi _{h_{\nu }}(\mu
_{\nu }|z_{\nu })\right\rangle _{(\lambda ,\tau )}=\frac{1}{Z_{(m,b)}}\int 
\mathcal{D}\phi \mathcal{D}^{2}\beta \mathcal{D}^{2}\gamma
\,e^{-S_{(m,b)}[\phi ,\beta ,\gamma ]}\prod_{\nu =1}^{n}\Phi _{h_{\nu }}(\mu
_{\nu }|z_{\nu })  \label{GH}
\end{equation}%
where the subscript $(\lambda ,\tau )$ on the left hand side stands to
remind of the functional measure $\int \mathcal{D}\phi \mathcal{D}^{2}\beta 
\mathcal{D}^{2}\gamma \,$ depending on the modular and the twist parameters.
The definition of correlation functions in (\ref{GH}) includes the genus-one
partition function $Z_{(m,b)}$, which we will discuss below.

As in the case of genus-zero calculation, the integration over the fields $%
\gamma $ and $\bar{\gamma}$ yield $\delta $-functions that fix the
conditions 
\begin{equation}
\bar{\partial}\beta (w)=2\pi \sum_{\nu =1}^{n}\mu _{\nu }\,\delta (w-z_{\nu
}),\ \ \ \ \ \partial \bar{\beta}(\bar{w})=-2\pi \sum_{\nu =1}^{n}\bar{\mu}%
_{\nu }\,\delta (\bar{w}-\bar{z}_{\nu }).  \label{dbeta}
\end{equation}

However, the solutions to (\ref{dbeta}) in this case is more complicated. To
integrate these equations on the torus it is convenient to introduce the $%
\theta $-function 
\begin{equation}
\theta (z|\tau )\ =\ -\sum_{n\in Z}e^{i\pi (n+1/2)^{2}\tau +2\pi
i(n+1/2)(z+1/2)},  \label{tita}
\end{equation}%
which obeys the periodic condition $\theta (z+p+q\tau |\tau )\ =\
(-1)^{p-q}\,e^{-i\pi q(2z+q\tau )}\theta (z|\tau )$, for $p,q\in \mathbb{Z}$%
. This property permits to build up from $\theta (z|\tau )$ a new function $%
\sigma _{\lambda }(z|\tau )$ which happens to have a single pole and the
same periodicity condition that we asked for $\beta $. That is, one can
define%
\begin{equation}
\sigma _{\lambda }(z,w|\tau )\ =\ \frac{\theta (\lambda +w-z)|\tau )\theta
\prime (0|\tau )}{\theta (z-w|\tau )\theta (\lambda |\tau )},
\end{equation}%
which, in fact, satisfies $\sigma _{\lambda }(z+p+q\tau ,w|\tau )\ =\
e^{2\pi iq\lambda }\sigma _{\lambda }(z,w|\tau )$. Then, one can use these
modular functions to integrate (\ref{dbeta}). The integration of these
equations is unique as long as the twist parameter $\lambda $ does not
vanish. We have 
\begin{equation}
\beta (w)=\sum_{\nu =1}^{n}\mu _{\nu }\sigma _{\lambda }(w,z_{\nu }|\tau )=u%
\frac{\prod_{i=1}^{n}\theta (w-y_{i}|\tau )}{\prod_{\nu =1}^{n}\theta
(w-z_{\nu }|\tau )},
\end{equation}%
where, again, a function $u$ appears. Field $\beta $ is a meromorphic
differential and depends on $n+1$ parameter; $n$ of these parameters are the
variables $\mu _{\nu }$ and the other parameter is $\lambda $. We can
parameterize $\beta $ in terms of $u$ and $n$ parameters $y_{i}$ by defining
the following set of $n+1$ implicit equations 
\begin{equation}
\mu _{\nu }=\frac{u\prod_{i=1}^{n}\theta (z_{\nu }-y_{i}|\tau )}{\theta
^{\prime }(0|\tau )\prod_{\mu \neq \nu ,\mu =1}^{n}\theta (z_{\nu }-z_{\mu
}|\tau )},\qquad \lambda =\sum_{\nu =1}^{n}(y_{\nu }-z_{\nu }),  \label{u}
\end{equation}%
where $\theta ^{\prime }(0|\tau )$ refers to the derivative of the $\theta $%
-function. Then, one has $n+1$ equations that relate variables $\mu _{\nu }$
and $\lambda $ with variables $y_{i}$ and $u$. This is exactly what has been
done in Ref. \cite{Schomerus} for the case $m=1$. Equation (\ref{u})\ comes
from the computation of the residue of the function $\beta $ at the pole $%
w=z_{\nu }$. Then, the integration over $\gamma $ and $\bar{\gamma}$ leads
to the following $\delta $-function 
\begin{equation}
\delta ^{(2)}(\bar{\partial}\beta (w)-2\pi \sum_{\nu =1}^{n}\mu _{\nu
}\delta ^{2}(w-z_{\nu }))\ =\ \left\vert \det \partial _{\lambda
}\right\vert ^{-2}\delta ^{(2)}\big(\beta (w)-u\mathcal{X}_{1}(y_{i},z_{\nu
};w)\big),  \label{delta}
\end{equation}%
where%
\begin{equation}
\mathcal{X}_{1}(y_{i},z_{\nu };w)\equiv \frac{\prod_{i=1}^{n}\theta
(w-y_{i}|\tau )}{\prod_{\nu =1}^{n}\theta (w-z_{\nu }|\tau )},  \label{X1}
\end{equation}%
and where the factor $\left\vert \det \partial _{\lambda }\right\vert ^{-2}$
is the Jacobian of the change of variables from $\partial \beta $ to $\beta $%
. Then, one can integrate over $\beta $ and $\bar{\beta}$ and finally find 
\begin{equation}
\left\langle \prod_{\nu =1}^{n}\Phi _{h_{\nu }}(\mu _{\nu }|z_{\nu
})\right\rangle _{(\lambda ,\tau )}=\frac{1}{Z_{(m,b)}|\det \partial
_{\lambda }|^{2}}\int \mathcal{D}\phi \,e^{-S_{\text{eff}}[\phi ,\mathcal{X}%
_{1}]}\prod_{\nu =1}^{n}\,|\mu _{\nu }|^{2m(j_{\nu }+1)}\,e^{2b(j_{\nu
}+1)\phi (z_{\nu })}
\end{equation}%
with the effective action 
\begin{equation*}
S_{\text{eff}}[\phi ,\mathcal{X}_{1}]=\frac{1}{2\pi }\int d^{2}w\left(
\partial \phi \bar{\partial}\phi +b^{2}|u|^{2m}|\mathcal{X}%
_{1}|^{2m}e^{2b\phi }\right) .
\end{equation*}

Shifting $\phi (w,\bar{w})\rightarrow \phi (w,\bar{w})-(m/b)\,\log |u|$ and
using (\ref{u}), one finds

\begin{equation}
\begin{split}
\left\langle \prod_{\nu =1}^{n}\Phi _{j_{\nu }}(\mu _{\nu }|z_{\nu
})\right\rangle _{(\lambda ,\tau )}& =\frac{1}{Z_{(m,b)}|\det \partial
_{\lambda }|^{2}}\int \mathcal{D}\phi \,e^{-S_{\text{eff}}[\phi ,\mathcal{X}%
_{1}]}\prod_{\nu =1}^{n}e^{2b(j_{\nu }+1)\phi (z_{\nu })}\times \\
& \times \prod_{\nu =1}^{n}\prod_{\mu \neq \nu }^{n}\left\vert \frac{\theta
(z_{\nu }-z_{\mu }|\tau )}{\theta ^{\prime }(0|\tau )}\right\vert
^{-2m(j_{\nu }+1)}\prod_{\nu =1}^{n}\prod_{i=1}^{n}\left\vert \frac{\theta
(z_{\nu }-y_{i}|\tau )}{\theta ^{\prime }(0|\tau )}\right\vert ^{2m(j_{\nu
}+1)}.
\end{split}%
\end{equation}

It is now convenient to define a new variable as 
\begin{equation}
\varphi (w,\bar{w})\equiv \phi (w,\bar{w})+\frac{m}{2b}\sum_{\nu
=1}^{n}\left( \log |\theta (w-y_{\nu })|^{2}-\log |\theta (w-z_{\nu
})|^{2}\right) ,
\end{equation}%
and introduce a new function $F$ defined as follows 
\begin{equation}
F(z-w|\tau )\ =\ e^{-2\pi (\text{Im}(z-w))^{2}/\text{Im}(\tau )}\left\vert 
\frac{\theta (z-w|\tau )}{\theta ^{\prime }(0|\tau )}\right\vert ^{2},
\label{d28}
\end{equation}%
which satisfies $F(z+p+q\tau -w|\tau )=F(z-w|\tau ).$ This permits to
rewrite the relation between $\varphi $ and $\phi $ in terms of single
valued variables, the fluctuation field $\phi _{\text{f}}=\phi +\phi _{\text{%
c}}$ and the function $F$. Namely, 
\begin{equation}
\varphi (w)\ =\ \phi _{\text{f}}(w)\,+\,\frac{m}{2b}\sum_{\nu =1}^{n}\left(
\log F(w-y_{\nu }|\tau )\,-\log F(w-z_{\nu }|\tau )\right) \,+\Delta
\end{equation}%
where $\Delta \ =\ (\pi m/$Im$(\tau )b)\sum_{\nu =1}^{n}\left( \text{Im}%
(y_{\nu })^{2}-\text{Im}(z_{\nu })^{2}\right) $. After some manipulation,
one finds

\begin{equation}
\begin{split}
\left\langle \prod_{\nu =1}^{n}\Phi _{h_{\nu }}(\mu _{\nu }|z_{\nu
})\right\rangle _{(\lambda ,\tau )}\ =\ & \ \frac{e^{-\pi \,m^{2}\text{Im}%
(\lambda )^{2}/\text{Im}(\tau )b^{2}}}{Z_{(m,b)}|\det \partial |_{\lambda
}^{2}}\ \frac{\prod_{\mu <\nu }^{n}F(z_{\mu }-z_{\nu }|\tau )^{\frac{m}{%
2b^{2}}}\prod_{i<j}^{n}F(y_{i}-y_{j}|\tau )^{\frac{m}{2b^{2}}}}{\prod_{\mu
=1}^{n}\prod_{i=1}^{n}F(z_{\mu }-y_{i}|\tau )^{\frac{m}{2b^{2}}}}\  \\
& \times \ \int \mathcal{D}\varphi \,e^{-S_{\text{L}}[\varphi ]}\prod_{\nu
=1}^{n}e^{2\alpha _{\nu }\varphi (z_{\nu })}\,\prod_{i=1}^{n}e^{-\frac{m}{b}%
\varphi (y_{i})}.
\end{split}
\label{ggg}
\end{equation}

On the left hand side of (\ref{ggg}) we already see the Liouville
correlation functions to appear. In order to normalize the correlation
functions we have to consider the partition function $Z_{(m,b)}$, which
depends on $\lambda $ and $\tau $. This function differs from the one for $%
m=1$ by a factor $e^{\pi \,(1-m^{2})(\text{Im}\lambda )^{2}/\text{Im}(\tau
){}b^{2}}$; \textit{c.f.} \cite{Schomerus}. The case $Z_{(m=1,b)}$
corresponds to the partition function of $\mathbb{H}_{+}^{3}$ WZW model. The
case $Z_{(m=0,b)}$ is, of course, the partition function of Liouville theory
, $Z_{\text{L}}$, times the contribution of the free $\gamma $-$\beta $
ghost system. Equation (\ref{delta}) imposes $\beta $ to be a constant,
which has be zero for $\lambda \neq 0$; in turn, the integration yields just 
$|\det \partial _{\lambda }|^{-2}$. In the untwisted case $\lambda =0$, $%
\beta $ may take any value and the integration diverges.

Collecting all the ingredients above, one arrives to the genus-one
generalization of the Ribault formula of \cite{Ribault-family}; namely%
\begin{equation}
\Omega _{(m,b)}^{g=1,n}=\left\langle \prod_{\nu =1}^{n}e^{i\frac{m}{b}%
X(z_{\nu })}\prod_{i=1}^{n}e^{-i\frac{m}{b}X(y_{i})}\right\rangle
_{X}\left\langle \prod_{\nu =1}^{n}V_{\alpha _{\nu }}(z_{\nu
})\prod_{i=1}^{n}V_{-\frac{m}{2b}}(y_{i})\right\rangle _{\text{L}},
\label{relation}
\end{equation}%
with the Liouville correlation function 
\begin{equation}
\left\langle \prod_{\nu =1}^{n}V_{\alpha _{\nu }}(z_{\nu
})\prod_{i=1}^{n}V_{-\frac{m}{2b}}(y_{i})\right\rangle _{\text{L}}=\frac{1}{%
Z_{\text{L}}}\int \mathcal{D}\varphi \,e^{-S_{\text{L}}[\varphi ]}\prod_{\nu
=1}^{n}e^{2\alpha _{\nu }\varphi (z_{\nu })}\prod_{i=1}^{n}e^{-\frac{m}{b}%
\varphi (y_{i})}
\end{equation}%
with exactly the same relation between the indices $j_{\nu }$ and the
Liouville momenta $\alpha _{\nu }=b(j_{\nu }+1+\frac{m}{2}b^{-2})$, but with
two additional degenerate fields inserted. The prefactor takes the form%
\begin{equation*}
\left\langle \prod_{\nu =1}^{n}e^{i\frac{m}{b}X(z_{\nu
})}\prod_{i=1}^{n}e^{-i\frac{m}{b}X(y_{i})}\right\rangle _{X}=\prod_{\mu
<\nu }^{n}F(z_{\mu }-z_{\nu }|\tau )^{\frac{m}{2b^{2}}%
}\prod_{i<j}^{n}F(y_{i}-y_{j}|\tau )^{\frac{m}{2b^{2}}}\prod_{\mu
,i=1}^{n}F(z_{\mu }-y_{i}|\tau )^{-\frac{m}{2b^{2}}},
\end{equation*}%
which, again, can be thought of as the expectation value of exponential
vertex operators in a theory of a free boson $X(z)$, now on the torus.

Relation (\ref{relation}) is valid for all values $m\in \mathbb{Z}_{\geq 0}$%
. It generalizes the genus-zero results of \cite{Ribault-family} to
genus-one, which has been accomplished by straightforwardly adopting the
analysis of \cite{Schomerus} to the generic case $m\in \mathbb{Z}_{\geq 1}$.
As a consequence, now we have (\ref{relation}), which relates the torus $n$%
-point function of the theory defined by action (\ref{accion}) to Liouville $%
2n$-point functions for $m\in \mathbb{Z}_{\geq 1}$.

\section{Conclusions}

Motivated by the question about how to extend the analysis of \cite{AT} to
the case of non-fundamental surface operators in $\mathcal{N}=2$ theories,
we reviewed the results of Ref. \cite{Schomerus} and used the path integral
techniques developed therein to generalize the result of Ref. \cite%
{Ribault-family} to genus-one. That is, we have shown that torus $n$-point
correlation functions of the conformal field theories proposed in \cite%
{Ribault-family}, whose Lagrangian representation is given by (\ref{accion}%
), are given by $2n$-point correlation functions of Liouville field theory
times a free field factor. In particular, this implies that the expectation
value of a surface operator of the $\mathcal{N}=2^{\ast }$ $SU(2)$ super
Yang-Mills theory whose type of singularity is labeled by an integer number $%
m$ is in correspondence with the expectation value of a primary operator of
one of the CFTs proposed in \cite{Ribault-family}, whose central charge is $%
c_{(b,m)}=3+6(b^{-1}+(1-m)b)^{2}$, with $b^{2}=\varepsilon _{1}/\varepsilon
_{2}$, and with $\varepsilon _{1,2}$ being the Nekrasov deformation
parameters. In the case $m=1$, which corresponds to the simplest surface
operators studied in \cite{AT}, the relation between correlation functions
mentioned above reduces to the WZNW-Liouville correspondence studied in
Refs. \cite{Stoyanovsky, Ribault-Teschner}, or, more precisely, to its
genus-one generalization done in Ref. \cite{Schomerus}. Our aim here was to
point out that the theories that correspond to other values of $m\in \mathbb{%
Z}_{>0}$ could also have applications to gauge theories through the AGT
conjecture and its generalizations. In particular, we have that the
Liouville two-point function that involves one degenerate field of level $m$%
, is given by correlator (\ref{relation}) in the case $n=1$, which in the
Coulomb gas representation takes the form of the following multiple integral
over the complex plane

\begin{equation}
\Omega _{(m,b)}^{g=1,n=1}\sim \Gamma (j+1)e^{2b(j+1)\delta (z,y)}\int
\prod_{k=1}^{-j-1}d^{2}\omega _{k}\prod_{k=1}^{-j-1}\frac{F(y-\omega
_{k}|\tau )^{m}e^{-2bm\phi _{\text{c}}(\omega _{k})}}{F(z-\omega _{k}|\tau
)^{2b^{2}(j+1)+m}\dprod\nolimits_{l\neq k}^{s}F(\omega _{k}-\omega _{l}|\tau
)^{2b^{2}}},
\end{equation}%
where $\delta (z,y)=\phi _{\text{c}}(z)+m(($Im$y)^{2}-($Im$z)^{2}+$Im$%
(z-y)^{2})/$Im$\tau $, $y=z+\lambda $.

The possible connection between the 2D CFT description of surface operators
in $\mathcal{N}=2$ $SU(2)$ gauge theories and the WZNW-Liouville
correspondence of \cite{Stoyanovsky, Ribault-Teschner} had been already
suggested in Refs. \cite{Gaston, AT}. A generalization to higher genus-$g$
worked out in \cite{Schomerus}, which can be easily extended to $m>1$ as we
did here for $g=1$, shows that $n$-point correlation functions of theories
with affine symmetry are given by ($2n+2g-2$)-point correlation functions in
Liouville theory, and the former are in correspondence with gauge theory
observables associated to having a surface operator for each trinion in the
Riemann surface decomposition. The question remains open as to how to make
the relation between the affine description of \cite{AT} and the
WZNW-Liouville correspondence of \cite{Stoyanovsky, Ribault-Teschner}
precise; if extended to generic $m$, it could provide a useful tool to
investigate non-fundamental surface operators. 
\begin{equation*}
\end{equation*}

This work has been supported by ANPCyT, CONICET, and UBA.

\end{document}